\documentstyle{article}
\begin{document}
\openup6pt

\title {Heckmann's wormholes in Jordan-Brans–Dicke gravity.}
\author{S.M.KOZYREV \thanks{Email address: Sergey@tnpko.ru}  \\
Scientific center gravity wave studies ''Dulkyn'', \\
PB 595, Kazan, 420111, Russia ,Kazan, Russian Federation \\}
\date{}
\maketitle

\begin{abstract}
A simple Heckmann's vacuum wormhole solution of Jordan-Brans-Dicke
gravitation is presented and analysed. It is shown that in
contrast with class I Brans solution where the throat radius
becomes real when $\omega < -4/3$ here it becomes positive when
$\omega < -1$.

\end{abstract}

PACS: 04.20.Jb, 04.40.Dg, 95.35.+d
\section{Introduction}
Lately there has been renewed interest in the scalar-tensor
theories of gravitation. Important arena where these theories have
found immense applications is the field of wormhole physics.
Several classes of solutions in the scalar-tensor theories support
wormhole geometry. The most prominent example of scalar-tensor
theories is perhaps the Jordan-Brans-Dicke (JBD) \cite{Jordan},
\cite{Brans}. Now a day, predictions of JBD theory to be
consistent not only with the weak field solar system tests but
also with the recent cosmological observations \cite{Nayak}.

    JBD theory describes gravitation through a metric tensor g$_{\mu \nu }$ and a massless
scalar field $\phi $. In this theory, static wormhole solutions
were found in vacuum, the source of gravity being the scalar
field. Several static wormhole solutions in JBD theory have been
widely investigated in the literature \cite{Agnese},\cite{Nandi},
\cite{Kim}. It was shown that three of the four Brans classes of
vacuum solutions admit a wormholelike spacetime for convenient
choices of their parameters.

In what follows, we shall present a static vacuum wormhole
solution of the BD theory endowed with first exact solution of JBD
field equations were obtained in parametric form by Heckmann
\cite{Heckmann}, soon after Jordan proposed scalar tensor theory.

\section{The Heckmann's wormholes solutions.}
JBD theory are described by the following action in the Jordan
frame is:
\begin{eqnarray}
S &=&\int dx\sqrt{-g}(\phi R-\omega  g^{\mu \nu }\nabla
_\mu \phi \nabla _\nu \phi -  \nonumber \\
&&\ \ \;\ \;\ \;\ \;\ \;\ \;\ \;\ \;\ \;\ \;\ \;\  )+S_m.
\label{eq3}
\end{eqnarray}

Here, \textit{R} is the Ricci scalar curvature with respect to the
space-time metric g$_{\mu \nu }$ and S$_m$ denote action of matter
fields. (We use units in which gravitational constant \textit{G}=1
and speed of light c=1.)

Variation of (\ref{eq3}) with respect to g$_{\mu \nu }$ and $\phi
$ gives, respectively, the field equations:

\begin{equation}
R_{\mu \nu }-\frac 12Rg_{\mu \nu }=\frac 1 {\phi \ }T_{\mu \nu
}^M+ T_{\mu \nu }^{JBD},  \label{eq4}
\end{equation}
where

\begin{eqnarray}
T_{\mu \nu }^{JBD} &=& [ \frac \omega {\phi ^2}\left( \nabla _\mu
\phi \nabla _\nu \phi -\frac 12g_{\mu \nu }\nabla _\alpha \phi
\nabla ^\alpha \phi \right) +  \nonumber \\[0.01in]
&&\ \ +\frac 1\phi \left( \nabla _\mu \nabla _\nu \phi -g_{\mu \nu
}\nabla _\alpha \nabla ^\alpha \phi \right) ].  \label{eq5}
\end{eqnarray}
and

\begin{equation}
\nabla _\alpha \nabla ^\alpha \phi =\frac{T_\lambda ^{M\ \lambda
}}{3 + 2 \omega },  \label{eq6}
\end{equation}
and $T_\lambda ^{M\ \lambda }$ is the energy momentum tensor of
ordinary matter which obeys the conservation equation $T_{\mu \nu ;\lambda }^{M\ }$ g$%
^{\nu \lambda }$= 0.

One can chose the static spherically symmetric metric in curvature
coordinates form
\begin{eqnarray}
ds^2=-e^{\nu \left( r\right) }dt^2+  e^{\lambda \left( r\right) }
dr^2+r^2d\Omega^2. \label{eq7}
\end{eqnarray}
 Then the solutions of the gravitational field
equations in the vacuum take the form \cite {Heckmann}
\begin{equation}
r=\frac{\alpha _0}{\sqrt{\tau }\left( \tau ^{-h}-\tau ^h\right)
},\nonumber
\end{equation}
\begin{equation}
e^{\lambda} =\frac{4h^2}{\left[ \left( \frac 12+h\right) \tau
^h-\left( \frac 12-h\right) \tau ^{-h}\right] ^2}, \nonumber
\end{equation}
\begin{equation}
e^{\nu} = \tau ^{\frac {1}{B}}, \label{e4}\\
\end{equation}
\begin{equation}
\phi =\phi _0\tau ^{\frac{\beta _0}B},\nonumber
\end{equation}
where $\tau $ parameter, arbitrary constant and

\[
h^2=\frac 14-\frac A{B^2};^{}A=\frac{\beta _0}2\left( 1-\beta
_0\omega \right) ;^{}B=1+2\beta _0.
\]
The constant $\phi _0$  are determined by an asymptotic flatness
condition as $\phi _0$ =1, while $\alpha_0$ and $\beta _0$ is
determined by the requirement of having Schwarzschild geometry in
the weak field limit \cite{Jordan}. Thereby
\begin{eqnarray}
\alpha_0 = 4 h M \left( 1+2 \beta _0 \right) \nonumber
\end{eqnarray}

is the function of central mass of the configuration and
\begin{eqnarray}
\beta _0 =- \frac{1}{3+2 \omega}. \label{eq191}
\end{eqnarray}
This implies that the range of $\beta _0$ is dictated by the range
of  $\omega$, which, in turn, is to be dictated by the
requirements of wormhole geometry.
 In order to investigate whether a given solution represents a
wormhole geometry, it is convenient to cast the metric into
Morris-Thorne canonical form:
\begin{eqnarray}
&& ds^2=-e^{2\chi \left( \stackrel{*}{R}\right) }dt^2+\left[
1-\frac{b\left( \stackrel{*}{R}\right) }{\stackrel{*}{R}}\right]
^{-1}dr^2+ \stackrel{*}{R}^2d\Omega ^2,\label{eq111}
\end{eqnarray}
where $\chi \left( \stackrel{*}{R}\right) $ and \textit{b}$\left(
\stackrel{*}{R}\right) $ are arbitrary functions of the radial
coordinate, $\stackrel{*}{R}$. $\chi \left( \stackrel{*}{R}\right)
$ is denoted as the redshift function, for it is related to the
gravitational redshift; \textit{b}$\left( \stackrel{*}{R}\right) $
is called the form function, because as can be shown by embedding
diagrams, it determines the shape of the wormhole \cite{Visser}.
The radial coordinate has a range that increases from a minimum
value at $\stackrel{*}{R}_0$, corresponding to the wormhole
throat, to infinity. The Heckmann solution can be cast to the form
(\ref{eq111}) by defining a radial coordinate $\stackrel{*}{R}$
which is related with \textit{r} via the expression
\begin{equation}
\stackrel{*}{R}=\frac{\alpha _0}{\sqrt{\tau }\left( \tau
^{-h}-\tau ^h\right) },
\end{equation}
The functions $\chi \left( \stackrel{*}{R}\right) $ and
\textit{b}$\left( \stackrel{*}{R}\right) $ are the given by

\begin{equation}
\chi \left( \stackrel{*}{R}\right) = \tau ^{\frac 1B},
\end{equation}
\begin{equation}
b\left( \stackrel{*}{R}\right) =\stackrel{*}{R}[ (
1- [\frac{ \left( 1+2 h\right) }{4h}\tau ^h  \nonumber \\
 - \frac{ \left( 1-2 h\right) }{4h}\tau ^{-h} ]^2 )
]
\end{equation}
The axially symmetric embedded surface $z=z \left(
\stackrel{*}{R}\right) $ shaping the wormhole's spatial geometry
is obtained from
\begin{equation}
\frac{dz}{d\stackrel{*}{R}}=\pm \left[ \frac
{\stackrel{*}{R}}{b\left( \stackrel{*}{R}\right) }-1\right]
^{-\frac 12}
\end{equation}
By definition of wormhole at throat its embedded surface is
vertical. The throat of the wormhole occurs at $\ \stackrel{*}{R}
= \stackrel{*}{R}_0$ such that $\ b\left( \stackrel{*}{R}_0\right)
= \stackrel{*}{R}_0$. This gives minimum allowed \textit{R} -
coordinate radii R$_0^{\pm }$ as \cite{Heckmann}
\begin{equation}
R_0^{\pm }= \frac {\alpha _0 \sqrt{ {1-4h}^2} }{4h} \left(
\frac{1+2h}{1-2h}\right)  ^ {  \frac{1} {4h}} \label{eqR}
\end{equation}
The values  R$_0^{\pm }$ can be obtained from (\ref{eqR}) using
the  (\ref{eq191}).
\begin{equation}
R_0^{\pm }= \sqrt{\frac{-3 \left( 1+\omega \right)}{2}} \left(
\frac{1+2\omega+\Omega}{1+2\omega -\Omega}\right) ^ {
\frac{1+2\omega} {\Omega}} \label{eqR},
\end{equation}
where $\Omega = \sqrt{7 + 10 \omega + 4 \omega^2}$.

 In contrast with class I Brans solution where the throat
radius becomes real when $\omega < -4/3$ here it becomes positive
when $\omega < -1$. This  range thus gives rise viable wormhole
geometry. The redshift function has a singularity at $\tau=0$
which corresponds to the point $\stackrel{*}{R}=0$.

\section{Discussion}
The field equations in JBD gravity theory are non linear in
nature. Moreover, the physical and the geometrical meaning of the
radial coordinate $r$ are not defined by symmetry reasons and are
unknown a priori. In this context, using the key assumption that
the Heckmann \cite{Heckmann} solution physically acceptable it was
constructed the  spherically symmetric wormholes.

Our analysis reveals that $\omega$ may take on arbitrary negative
values less then -1. This result extends the scope for the
feasibility of wormhole scenarios even to the regime of ordinary
observations.

\noindent {\bf Acknowledgments} The author is grateful to R.A.
Daishev and S.V. Sushkov for the useful discussions. The work was
supported in part by the Kazan Institute of Applied Problems.
 \\

\end{document}